# Exploring Micro-Services for Enhancing Internet QoS

Deval Bhamare
Qatar University,
Doha, Qatar
devalb@qu.edu.qa

Mohammed Samaka
Qatar University,
Doha, Qatar
samaka.m@qu.edu.qa

Aiman Erbad
Qatar University,
Doha, Qatar
aerbad@qu.edu.qa

Raj Jain
Washington Univ.,
St. Louis, USA
jain@wustl.edu

Lav Gupta
Washington Univ.,
St. Louis, USA
lavgupta@wustl.edu

*Abstract:* With the enhancements in the field of software-defined networking and virtualization technologies, novel networking paradigms such as network function virtualization (NFV) and the Internet of things (IoT) are rapidly gaining ground. Development of IoT as well as 5G networks and explosion in online services has resulted in an exponential growth of devices connected to the network. As a result, application service providers (ASPs) and Internet service providers (ISPs) are being confronted with the unprecedented challenge of accommodating increasing service and traffic demands from the geographically distributed users. To tackle this problem, many ASPs and ISPs, such as Netflix, Facebook, AT&T and others are increasingly adopting micro-services (MS) application architecture. Despite the success of MS in the industry, there is no specific standard or research work for service providers as guidelines, especially from the perspective of basic micro-service operations. In this work, we aim to bridge this gap between industry and academia and discuss different micro-service deployment, discovery and communication options for service providers as a means to forming complete service chains. In addition, we address the problem of scheduling micro-services across multiple clouds, including micro-clouds. We consider different user-level SLAs, such as latency and cost, while scheduling such services. We aim to reduce overall turnaround time as well as costs for the deployment of complete end-to-end service. In this work, we present a novel affinity-based fair weighted scheduling heuristic to solve this problem. We also compare the results of proposed solution with standard greedy scheduling algorithms presented in the literature and observe significant improvements.[1]

## I. Introduction

With the explosion of online services as well as mobile and sensory devices, demand for new services and consequently data traffic is growing rapidly. The popularity of Internet of Things (IoT) and 5G networks have contributed significantly to this trend, with millions of new sensing devices and online services exchanging data. According to Wireless World Research Forum (WWRF), the number of connected wireless devices is expected to be 100 billion by 2025 [1]. Cloud computing has been considered as a major enabler for such novel networking paradigms [2]. The online services, sensing devices, as well as end-users, are generally spread across geographically distributed areas. This motivates the ASPs and ISPs to deploy the services over multiple clouds for scalability, redundancy and quicker response to the users [1, 3, 4]. Increasing use of virtualization technologies also helps ASPs and ISPs to deploy their services over standard high-volume infrastructures to accommodate such high volume of user demands.

Large services, which were monolithic software in the past, are being replaced by a set of lightweight services called micro-services [5, 6], which are being deployed in distributed virtualized environments. Monolithic applications are complex, hard to scale, difficult to upgrade and innovate. On the contrary, micro-services, where the functionalities of the application are segregated, are lightweight, easy to deploy and scale. Instead of building a single, monolithic application, the idea is to split the application into a set of smaller, interconnected services, called micro-services (or simply services) [7]. Such services are lightweight and perform distinct tasks independent of each other. Hence, they can be deployed quickly and independently as user demands vary. The micro-services can be easily upgraded or scaled without affecting much of the other functionality. Spreading micro-services across multiple clouds allows having ASP's points-of-presence close to the distributed mobile users. The services are then chained through a process called service function chaining (SFC) [8] to create a complete end-to-end service. The goal is to enable the traffic to flow smoothly through the network, resulting in an optimal quality of experience to the users.

Micro-services can be easily deployed over physical machines (PMs) or virtual machines (VMs) using novel techniques such as containers, allowing service providers to easily deploy, scale and load balance their applications [5, 6, 9]. Micro-services deployed using containers benefit from lower maintenance, lower costs and more scalability as compared to directly deploying the virtual machines. The ASPs such as Google, Netflix, and others, are already using containers extensively. ISPs like AT&T and international communities like IETF are actively proposing the use of micro-services to bolster SDN and NFV efficiency [12, 13, 14]. For example, AT&T has started an "*open container initiative*" [13]. The users benefit from the quick response and lower costs, while ASPs and ISPs benefit from quicker and cheaper deployment options. Micro-services are usually scaled dynamically depending on the user demands. Many service providers are opting for micro-services to deploy their services. With the advancements in the virtualization technology, the micro-services are being deployed over virtual machines (VMs) as well. ASPs and ISPs send requests to cloud service providers (CSPs) and obtain the resources to deploy the micro-services as per their requirements at the time [3]. We discuss more about micro-services in Section III.

---
[1] This is an extended version of a paper presented at IEEE ICC 2017. We have extended the work significantly (50% changes) for submission to Journal of Transactions on Emerging Telecommunications Technologies (ETT).

Another recent trend is the movement of micro-services from host-centric to data-centric model in which the computational resources move closer to the end users. This results in further reduction in response time to the end-users and lower costs to ASPs and ISPs because of shorter access links. This has led service providers to the concept of micro-clouds at the cellular base stations [3, 13]. The technology is called as Micro Edge Computing or MEC. A sample scenario is demonstrated in Fig. 1. We consider the example of *Netflix*, a multinational entertainment content provider, which specializes in streaming media and video on demand. As a result, of an explosion in mobile devices [1], *Netflix* would benefit from locally relevant content cached at micro-clouds served to users through a micro-service. This will reduce the user-latencies and result in better user experience. In addition, it may result in lower operational expense (OpEx) to *Netflix* by reducing the usage of expensive wide area network (WAN) bandwidth.

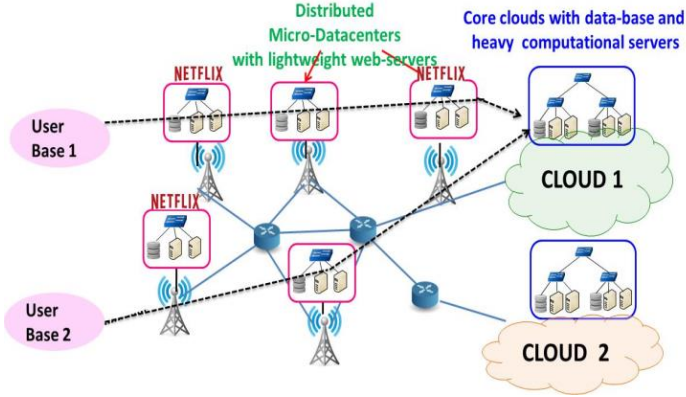

Fig. 1. Micro-Clouds at the base station for quicker response.

Due to the nature of the contemporary telecommunications applications, the services need to be highly available, almost as much as 99.999% [13]. Additionally, most of the contemporary applications are sensitive to the delays, jitter, and packet-loss (such as online games, healthcare applications, video streaming and others). Many of these services are required to support millions of subscribers and meet the rigorous performance standards [13, 16]. Proper scheduling of these services is important for reducing total delays, total required resources, and overall deployment costs [14]. These requirements mandate the optimal placement and scheduling of the service instances and proper interconnection among them.

Although virtual machine placement problem has already been studied in the literature [9-11, 44-48], micro-service architecture and its scheduling is a relatively novel problem. The instances of the micro-services are generally short-lived and dynamic in nature. Researchers are working on innovative schemes to design efficient algorithms for appropriately placing as well as scheduling the services [8, 13], splitting the load across instances on multiple clouds, and chaining them to improve performance parameters. However, we argue that there is a lack of research work in the domain of micro-service scheduling across multiple clouds for optimal service function chains (SFCs), for both ASPs and ISPs [6].

Despite the widespread acceptance of the micro-services in the industry, there is a huge gap between industry and academia in this field [25-28]. Hence, in this work, we aim to bridge this gap by discussing the micro-services and some of the options for deployment and discovery of micro-services as well communication among different instances of micro-service. In addition, we formally discuss the problem of scheduling micro-services. The service providers may benefit from such work to understand the limitations, challenges, and advantages of this novel networking architecture. Such work may motivate ASPs and ISPs to leverage advantages of micro-service and multi-cloud platforms.

The rest of the paper is organized as follows. In the next section, we discuss the state-of-the-art of micro-services and the scheduling problem in the SFC context to show the limitations of existing approaches. In Section III, we discuss the options for deployment and discovery of micro-services as well as communication among different instances of the micro-services to form end-to-end service chains. Section IV formalizes the micro-service scheduling problem. In Section V, we propose a novel *FWS* algorithm for micro-service scheduling and explain the experimental setup, and in Section VI, we present the comparison results. Finally, Section VII concludes the paper. A list of the acronyms used throughout this paper is given in Table 1.

Table I. List Of Acronyms

| *Acronym* | *Description* |
|---|---|
| API | Application program interface |
| ASP | Application service provider |
| CAPEX | Capital expenditures |
| CSP | Cloud service providers |
| DB | Database |
| DPI | Deep packet inspector |
| EC2 | Elastic Compute 2 |
| ESB | Enterprise service bus |
| FWS | Fair weighted affinity-based scheduling |
| IaaS | Infrastructure as a service |
| IoT | Internet of Things |
| ISP | Internet service provider |
| IT | Information Technology |
| LFDT | Least-full first with decreasing time |
| LFFF | Least-full first with first finish |
| MFDT | Most-full first with decreasing finish |
| MFFF | Most-full first with first time |
| MORSA | Multi-objective resource scheduling |
| MS | Micro-service |
| NFV | Network function virtualization |
| OF | OpenFlow |
| OPEX | Operational expenses |
| OSGi | Open Service Gateway Initiative |
| PM | Physical machine |
| REST | Representational State Transfer |
| SDN | Software-defined networking |
| SLA | Service level agreement |
| SFC | Service function chaining |
| SOA | Service-oriented architecture |
| VF | Virtual function |
| VM | Virtual machine |
| VNF | Virtual network function |
| WAN | Wide area network |
| WWRF | Wireless World Research Forum |

## II. Related Work

Micro-services are being extensively used in the industry. However, research community lacks the intensity with respect to various micro-service aspects. There are only a few works, e.g. [31-36], that discuss the micro-services. Most of the works on micro-services are available on blogs and online communities, mostly in a scattered manner [5, 42]. Johannes et al. and Dmitry et al. discuss the micro-service platform architectures in brief in [6, 11], respectively. Balalaie et al. provide the micro-service architecture for clouds in [14]. Newman discusses different options for building micro-services in [25]. Garderen also discusses micro-service architecture in brief [29].

Recently researchers have been studying SDN and NFV integration with micro-services as well. Authors in [49] leverage the NFV and micro-service architectural style to propose an architecture for on-the-fly CDN component provisioning to tackle issues such as flash crowds. In the proposed architecture, CDN components are designed as sets of micro-services which interact via RESTFul Web services and are provisioned as Virtual Network Functions (VNFs), which are deployed and orchestrated on-the-fly. Luong et al. [50] present a micro-service platform to target the flexibility of telecom networks and the automation of its deployment. Using Docker orchestration, this demo paper shows the flexibility and the rapid deployment of wireless network infrastructure. In [51], authors introduce tunable and scalable mechanisms that provide NFV MANO with high availability and fault recovery using micro-services. Fazio et al. [52] argue that developers can engineer applications that are composed of multiple lightweight, self-contained, and portable runtime components deployed across a large number of geo-distributed servers and discuss open issues in micro-service scheduling. Authors in [53] and [54] investigate further into fog-computing to reduce the latencies in LTE and 5G networks respectively, where micro-services can be a candidate solution for service deployment. Yaseen et al. [55] and Brito et al. [56] discuss leveraging fog computing and SDN for the security of the mobile wireless sensor networks and smart factories.

Micro-service architecture is being implemented by large enterprises such as banks, financial institutions, global retail stores and others to build their services in an incremental, flexible and cost-effective manner. Recently, there has been a trend to use containers to deploy micro-services across geographically distributed clouds. Containers are the lightweight version of the virtual machines. They are gaining significant traction in the industry recently since they are lightweight as compared to VMs. They can be easily downloaded and quickly deployed [29, 30]. Platforms such as Docker, Solaris Zones [33, 41] are available to ASPs for deployment of their services using containers [15]. Researchers have identified the importance of micro-services as an enabler for novel networking paradigms such as IoT and 5G. They have started identifying and addressing various problems in this context [5, 6].

One important problem in the context of micro-services is their scheduling over the available and scattered resources to form complete service function chains. Optimal scheduling of micro-services is necessary for faster deployments and minimum expenses to the service providers as well the better quality of experience to the end-users, such as lower latencies. The problem of placing and scheduling the virtual functions has been actively pursued in the industry and academia for years. However, researchers argue that the problem needs to be revisited from the perspective of micro-services over service chains, as service function chains (SFCs) has some unique features [18]. For example, SFC is an ordered chain of services, so the order in which the service instance needs to be visited is defined dynamically by the traffic flows [7]. The SFC scheduling problem has been in focus recently and works such as in [8-12, 16-21] provide a wide range of VM scheduling strategies in a single cloud or across multiple clouds forming efficient SFCs.

Due to the time-sensitive nature of contemporary applications, VM placement alone is not sufficient to yield acceptable performance in the deployment of micro-services over micro-clouds. Especially from the perspective of the short-lived micro-services, scheduling is more important than the placement problem. Also, mobile users have strict SLAs as far as tariffs and delays are concerned. This mandates ASPs to create points of presence close to the mobile users, reducing access latency and the overall cost. Merely efficiently placing the micro-services is not sufficient to obtain optimal results. Recently, researchers have become aware of the importance of scheduling problem for micro-services in SFCs, especially for the micro-clouds at the edges to guarantee carrier-grade performance [18]. However there is a dearth of research works which address the scheduling problem in the context of micro-services.

In this work, we propose a novel fair weighted affinity-based scheme for scheduling micro-services and compare results with four different variants of the greedy strategies, which are common in the literature. We show significant improvements with the proposed heuristic. In the next section, we discuss the options for deployment, the discovery of micro-services as well communication among the different instances of the micro-services.

## III. Micro-services and Options

As defined in [25, 47], the micro-service architecture is a specialization of an implementation approach for service-oriented architectures (SOA) used to build flexible, independently deployable software systems. Generally, software applications become easier to build and maintain when they are divided into smaller pieces, which cooperate to perform one particular complex task. For this discussion, we consider the example of an ASP who provides e-commerce based services. This may apply to ASPs such as Netflix, Uber, Amazon and many others who provide their services through the Internet. A single user request for some online service may comprise of a set of functionalities, which are accounting, storage, inventory and shipping [5, 42]. Each such functionality may be deployed as a micro-service (Fig. 2). The point to be noted here is that the scope of micro-service architecture is not only limited to the ASPs but also is equally important for the transport services, multimedia services as well as network services [13].

Micro-services communicate with each other to provide the desired functionality to end-users. For example, the user request for online service or product has to travel through the micro-service managing inventory, then accounting, storage and finally shipping to complete the order. Micro-services perform the allocated task and then exchange the related data to update the other micro-services. For example, in this case, inventory micro-service will update the accounting micro-service regarding total quantities ordered of the goods or services to prepare the bill. The same task may be performed using traditional monolithic services or even virtual machines. However, micro-services are more agile, lightweight and easy to scale up or down as per the user demands vary. In the remaining section, we discuss various ways to deploy different ways in which micro-services can be deployed to form complete service chains for an ASP's services.

1. *Micro-service Deployment*: With the advent of the virtualization technology, many micro-service deployment options are becoming available to the service providers. Below we discuss these options and the pros and cons associated with each one.

    a. *Multiple Service Instances per Host*: This is the simplest way of micro-service deployment, where multiple micro-service instances are deployed on a single host. Though the host could be a physical machine or a virtual machine, physical machines are preferred in this simpler way of deployment. This scheme benefits from the high resource utilization. However, it suffers from some significant drawbacks. The major drawback is that there is little isolation for different service instances. A single service instance may consume a significant amount of resources starving other service instances. In addition, security becomes a major threat in such an environment. This option is shown in Fig. 2.

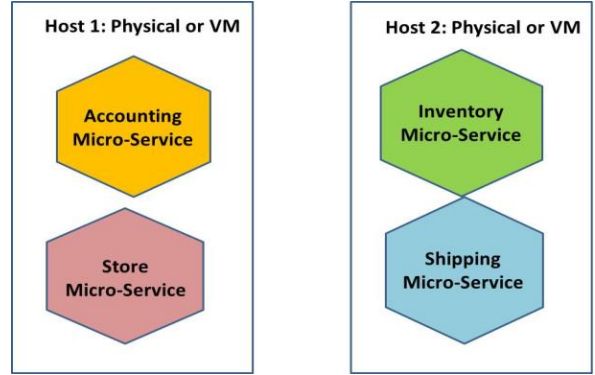

Fig. 2. Multiple service instances per host (generally PM).

   b. *Single Service Instance per Host*: In this approach, a separate host is selected for each micro-service. Generally, virtual machines (VMs) are selected as hosts. VM types are selected as per the system requirements of the service. Multiple hosts such VMs are then deployed over a single or multiple servers as per the capacity and other constraints. Recently, this is the primary approach used by a majority of the ASPs. Cloud service providers (CSPs), such as Amazon, make the resources available through the Infrastructure as a Service (IaaS) model. The benefit of this approach is that each service instance is executed within a completely isolated environment. It has a fixed amount of CPU and memory and does not have to share resources with other micro-services. With this model, ASPs can leverage mature cloud infrastructure. Single service instance per VM deployment is shown in Fig. 3 with account service as an example. However, since VMs are available in fixed sizes, it is possible that some VMs will be underutilized. Also, shutting down or restarting a particular VM instance, as user demands vary, is time-consuming and may affect the user-latencies adversely.

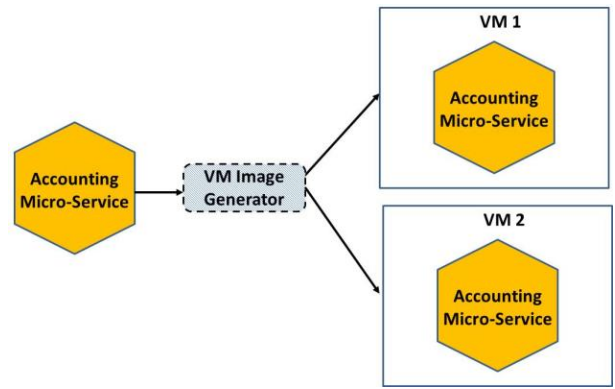

Fig. 3. Single service instance per host (generally VM)

   c. *Single Service Instance per Container*: Containers are the lightweight version of the virtual machines. A container is an OS-level virtualization to deploy and run applications without launching entire VM and is suitable for the deployment of the micro-services. A container consists of the libraries required to run the entire micro-service. A container may host multiple micro-services. More details about the containers may be found in the works such as [27, 35, 36] and online resources such as [5, 42]. Containers are gaining significant traction in the industry recently since they are lightweight as compared to VMs. They can be easily downloaded and quickly deployed. Containers may be deployed over physical machines or virtual machines quickly as per the choice of the service providers. The platforms such as *Docker* are generally used for the deployment of containers [33]. Micro-service deployment option using containers is shown in Fig. 4.

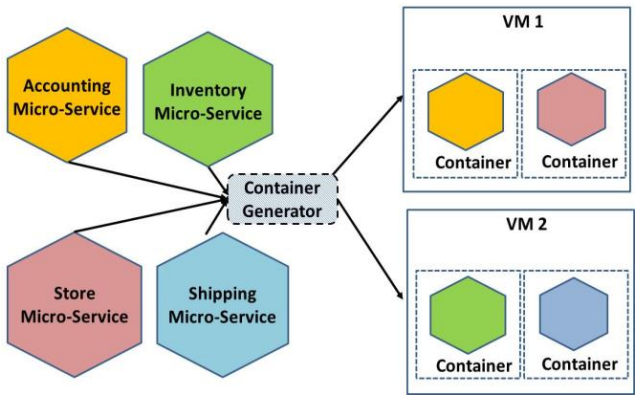

Fig. 4. Single service instance per container.

2. ***Discovery of Micro-services***: Once the micro-services are deployed, next important stage is the discovery of such services for the end-users, so that the user requests can be guided through the underlying network properly. In a service-oriented architecture (SOA), the inter-service communication among the service instances and service discovery module are implemented with an Enterprise Service Bus (*ESB*) [37]. For the efficient discovery of micro-services, various platforms such as Kubernetes and Marathon [42], which implement the service registry, have been developed in the industry. A service registry is nothing but a database of available service instances. Service registry module translates user requests into the appropriate message types and routes them to the appropriate provider, by enabling users to interconnect with the different services.

Examples of the service registry are *Apache Zookeeper, Netflix Eureka,* and others. The *OSGi* architecture [38] also provides a similar platform for service registry, discovery, and deployment of the services. These specifications enable a development model where applications are composed of many different reusable components. Contemporary service discovery architectures can be divided into two types, which are discussed below.

a. *Client-side Discovery*: In this approach, the client or the API-GW is responsible for obtaining the location of a service instance by querying a service registry. The user is also responsible for load balancing among the service instances. A sample client-side discovery model is shown in Fig. 5. We assume that the micro-service instances implement Representational State Transfer (REST) APIs [39, 40] for the communication purpose. The service provider is responsible for implementing the service registry; however, it is the responsibility of the clients or end-users to determine the location of the micro-service instance from the service registry. *Netflix OSS* is a good example of client-side discovery model [42]. Though this model is simple to implement, it couples the client code with the service registry.

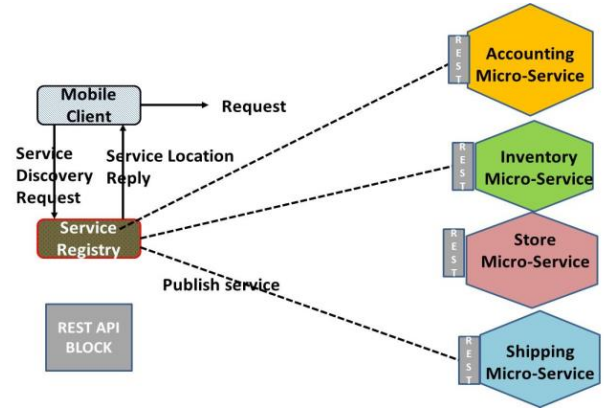

Fig. 5. Client-side service discovery.

b. *Server-side Discovery*: With this approach, clients/API-GWs send the request to a component, such as a load balancer, that runs in a well-known location. The load balancer is responsible for calling the service registry and determining the absolute location of the micro-service, as shown in Fig. 6. That component has an entry for the port and IP address for the service registry. Amazon web service elastic load balancer (*AWS ELB*) [43] is a good example of server-side discovery. The major advantage of this model is that the details of service registry are abstracted from the client.

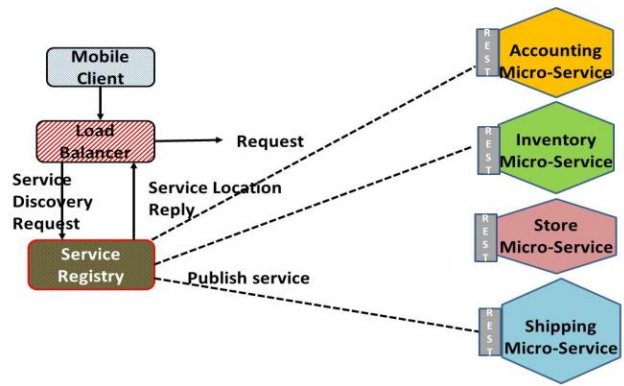

Fig. 6. Server-side service discovery.

3. ***Micro-service Communication***: In monolithic applications, different components invoke one another using language-level function calls. In contrast, in micro-service based applications, each service instance is typically a process. There are synchronous and asynchronous modes of communication among processes and micro-services use combination of these interaction styles. Communication among various micro-service instances is important for a complete service to the end users. There should be a proper mechanism to guide the user packets through proper instances of the micro-services in the given order [45]. For example, with the given hypothetical ASP in this work, the order of service instances through which the user request should be routed is: (*inventory→account→shipping→store*). From the design perspective, following communication options are available for the ASPs in their micro-service setup.

   a. *Point to point*: This is the simplest approach in which each service instance directly communicates with another using the APIs such as REST. The user directly communicates with the first service instance as shown in Fig. 7. As the number of functionalities and instances of each sub-service increase, the performance of systems based on point-to-point communication degrades. In addition, it gets unmanageable with a large number of service instances.

   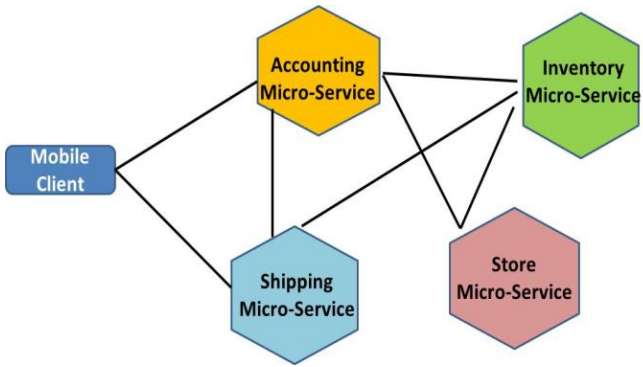

   Fig. 7. Point-to-point micro-service communication.

   b. *Communication through the API gateway*: In this option, an application program interface gateway (API-GW) is installed in-between the end-users and the micro-service instances, as shown in Fig. 8. API-GW has a well-known port-IP combination, to which, clients send the requests and API-GW forwards the requests to the appropriate service instance. The decision is taken based on parameters such as possible delay, load balancing, and others. Service instances do not have to bother with the communication among each other, and they just forward their replies to the gateways. This is more scalable as compared to the previous option. However, a single gateway may become a single point of failure. This disadvantage can be easily eliminated by having multiple instances of such gateways for load balancing and redundancy (Fig. 8).

   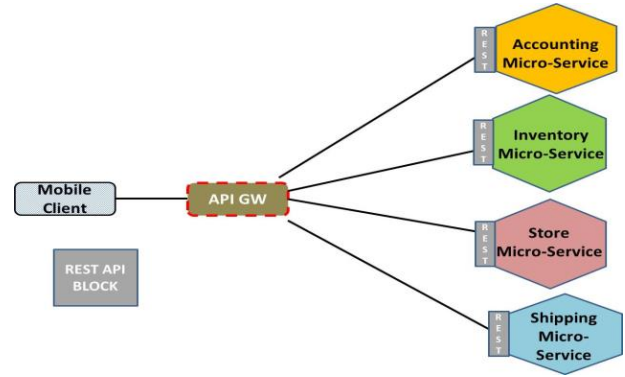

   Fig. 8. Communication through API-gateway.

   c. *Message Broker Style*: This is an asynchronous mode of communication. A given micro-service can be a message producer and can asynchronously send messages to a queue. On the contrary, the consuming micro-service takes messages from the queue. Such style of communication decouples message producers from message consumers and the intermediate message broker buffers messages until the consumer is able to consume or process them. Producer micro-services are completely unaware of the consumer micro-services, hence, are said to be in asynchronous communication [26, 27, 42]. Micro-service communication using message broker style is shown in Fig. 9.

   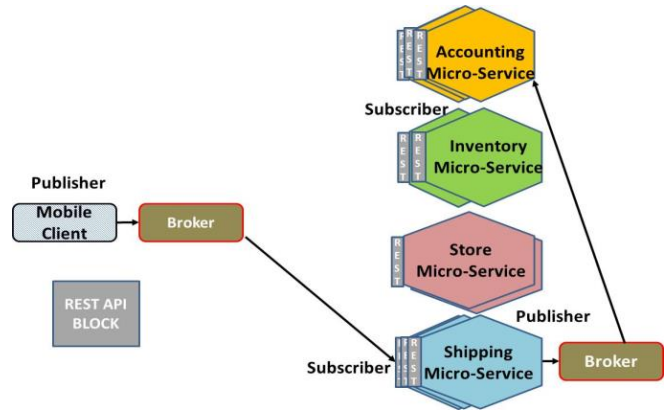

   Fig. 9. Message-Broker Style Communication option.

In the next section, we discuss one particular and important problem in the context of micro-services, that is, scheduling of micro-services. This is an important problem for optimal placement of service chains. Optimal scheduling is necessary to satisfy the service level agreements (SLAs) and QoS to end-users (such as minimum latencies) as well as minimum expenses to the service providers while deploying the instances of micro-services over the available resources. As pointed out earlier, though the problem of placement of virtual machines is studied quite extensively in the literature, scheduling of micro-services along with relevant micro-service options are severely under-researched.

## IV. Micro-services Scheduling Problem

In this section, we discuss the micro-service scheduling problem in the context of constitution and placement of SFCs. The problem generally comprises three sub-problems: (1) selecting types and numbers of the service instances to be scheduled (2) selecting physical machines (PMs) or virtual machines (VMs) on which the services should be scheduled and (3) deciding the time slot for which a particular service instance needs to be executed. Common heuristics used in the state-of-art systems for these tasks are "*greedy with bias*" [8, 19, 35]. The bias is towards some factor such as: (1) select a service with earliest finish time or (2) select service with the longest execution time. Similarly, the bias while selecting VMs/PMs are: (1) select most-loaded machine or (2) select least-loaded machine [11, 16]. We start our discussion with a particular use case. We consider an ASP such as *Facebook* (*FB*) and take up a hypothetical example of the services offered by FB to explain the problem under consideration. It is important to note that the scope of the problem under consideration is not only limited to the application services, but is equally important for the telecommunication services, multimedia services, and network services as well [7, 8]. For example, the network-slicing problem for the network service providers [13, 49].

As shown in Fig. 10, different groups of users from various user-bases may send different types of web requests to *FB* webserver(s). For example, some users may be interested in signing up for the service and others may log in to check their messages or posts on the wall or scan through their friend requests. The sign-up requests, after passing through the firewall, are passed to a set of services, which handle user registration logic (in this case *firewall* → $f_1$ → $f_2$ → $f_3$ → $f_4$ → $f_5$ → *database*). However, login requests may have to be passed through deep packet inspection (DPI) in addition to the firewall to distinguish among user demands (such as wall-post, photo upload or online *FB* integrated games). A complete service chain may comprise of a combination of IT and telecommunication services. This example is just for an illustration purpose of the service flows, and it may be different in actual *FB* implementation of the services. The important point to be noted here is the dynamic formation of complex and hybrid service chains, which comprise a different set of micro-services implemented at the application layer. Some of the long-lived services in the process such as the firewall, the deep packet inspection (DPI) and the database (DB) may stay for longer durations compared to other short-lived services such as function specific micro-services. Whereas placement would be enough for such long-lived services, the short-lived micro-services needs to be scheduled for efficient service chains [49].

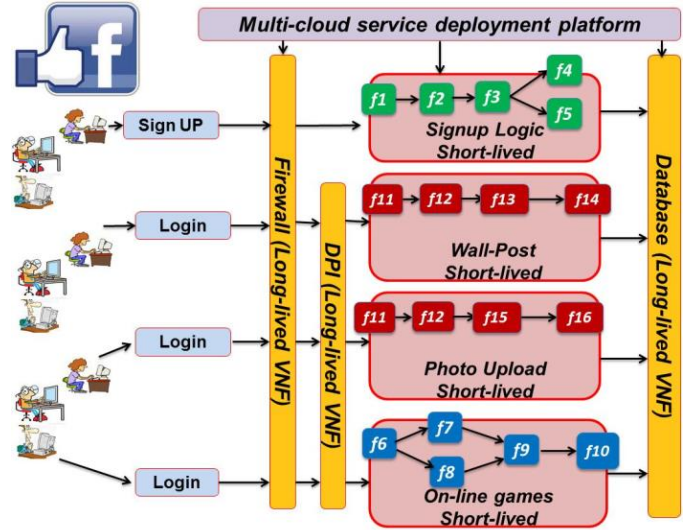

Fig. 10. SFCs for different services offered by an ASP (such as *FB*).

In this example, if we consider some specific functionality, such as user registration (sign-up), wall-post on *FB* or other integrated game applications, a specific set of service instances need to be executed. Such sets of service instances may be switched on/off as user demands vary, especially at the micro-clouds, since the capacities are limited. Scheduling these service instances over the available resources is an important problem. In this work, we have considered four SFCs comprising twenty micro-services in total. The SFC shapes and graphs are shown in the Fig. 11. Note that the topologies of the SFCs also indicate their execution order. For example, in SFC 1, service $f_2$ has to be executed after $f_1$. This may be because of the business logic dependence or some mandatory network traffic flow demand. For example, web-service logic handling service has to be executed before the service handling databases; or firewall must be executed before the business logic, etc.

However, $f_4$ and $f_5$ may be executed in parallel after $f_3$, since they are independent of each other. Similarly, in SFC 2, $f_7$ and $f_8$ may be executed at the same time after $f_6$. However, $f_9$ has to be executed only after both $f_7$ and $f_8$ have finished their execution. This mandatory ordered flow of services in SFCs makes scheduling a complex problem. For the sake of simplicity, we assume that the VFs are visited in the numerical order. There may exist different service flows following different chains. However, the numbers for the VFs are in numerical order. For example, different chains consisting of different VFs may exist, such as (1,2,3,4), (1,2,3,5), (6,7,9,10), (15, 17, 20) as shown in Fig. 11.

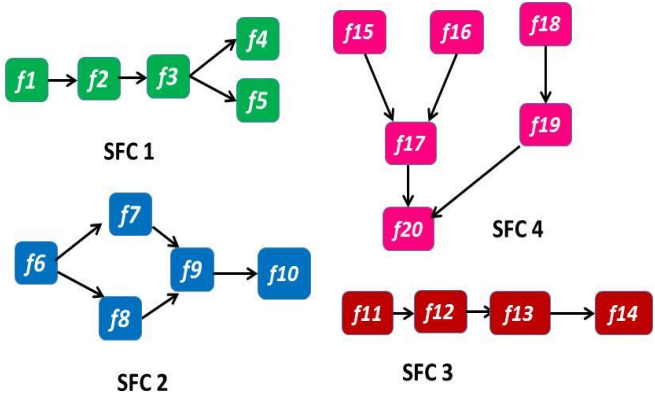

Fig. 11. Four SFCs with 20 virtual functions (VFs) used for evaluation.

Let us now consider the scheduling problem of micro-service by considering three SFCs from the example above displayed on the left of Fig. 12. On the right-hand side, we show the Gantt chart for the scheduling of the micro-services over available resources, using the virtual machines ($VM_1$ to $VM_5$), deployed across three clouds $C_1$, $C_2$, and $C_3$. Vertical lines indicate the time slots and each service needs different time to finish the execution. We assume that three user requests for these three SFCs arrive at the same time. The widths of the micro-services indicate the total time needed to execute the services (longer services mean longer time for execution). A possible scheduling to optimize the total time and the resources required for the three SFCs on the available resources is shown in Fig. 12.

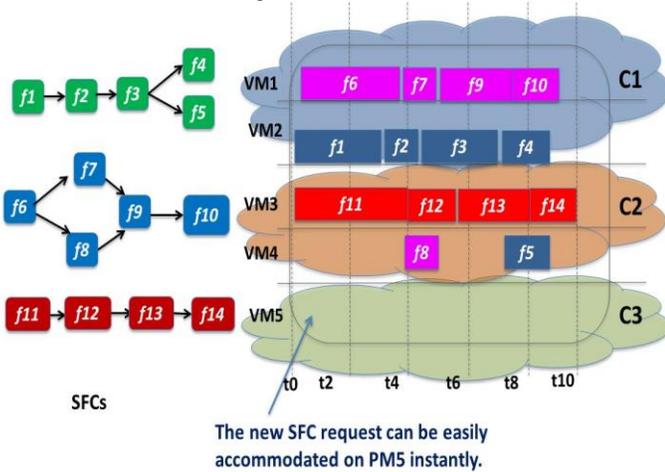

Fig. 12. Gantt chart for optimal scheduling.

We observe that, with optimal scheduling, all the executions finish before time-slot $t_{10}$ keeping $VM_5$ free and ready to serve another incoming request. We argue that a sophisticated heuristic is needed to solve the large-scale micro-service scheduling problem within acceptable time limits. In the next section, we propose our novel affinity-based fair weighted scheduling (FWS) scheme and explain the experimental setup. We implement all the four combinations of SFCs along with our proposed FWS approach. Our proposed novel heuristic performs scheduling of micro-services on multiple VMs/PMs spread across multiple clouds. We consider different user-level service level agreements (SLAs), such as traffic-affinity among services [8], user delays, and cost constraints. In addition, we consider network parameters such as link loads and network traffic. We aim to reduce the overall turnaround time for the service and reduce the total inter-VM traffic generated.

## V. Heuristics and Experimental Setup

In this section, we propose a novel affinity-based fair weighted scheme (FWS) for the scheduling problem under consideration. The heuristic can be divided into two distinct parts, that is, (1) selection of next service instance to be scheduled and (2) selection of next machine (VM or PM) on which the service instance should be scheduled. Heuristic starts at the time $t = t_0$. User requests arrive dynamically with inter-arrival time exponentially distributed, that is, the arrival process is *Poisson* [16, 23]. Let $U$ be the set of users, waiting for the service or being served at any time $t$. Initially, we prepare the graphs for each SFC for each user $u$ in $U$. It is to be noted that the graph may have disjoint sets of sub-graphs.

A sample inline service graph is shown in Fig. 13. Solid, dotted and dashed lines highlight three possible service chains (there may be several other SFCs as well). Also, the users may demand a single functionality, such as $F_9$ shown in the figure. Again, these graphs can be of any shape and size, depending on the service provided by a specific ASP and the types of end-user demands. We have used various resource combinations (from Amazon EC2 [22]) mentioned in Table II to simplify configurations so that resource requirements can be easily mapped to the nearest available configuration. Depending on the user resource demands, a particular VM is chosen from Table II such that the requirements are the closest match. Initially, we assign labels to the services using Coffman-Graham algorithm [9].

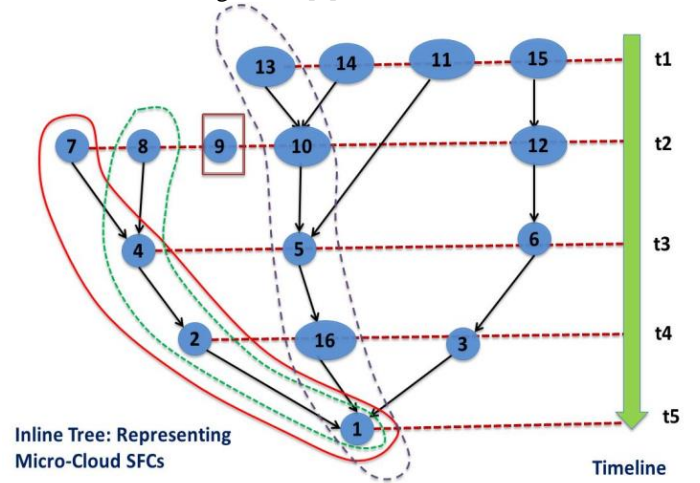

Fig. 13. An inline graph for services forming different SFCs.

It ensures that the service instance that needs to be executed first for the particular SFC (starting service) gets a priority as per the arrival time. The service instance with the highest value of the label is scheduled first. Further, we assign weight $w$ to the services, such that:

$w\ \alpha$ (number of dependent services in that chain) and
$w\ \alpha$ (time spent by the services in the waiting queue).

If there are ties between two services for scheduling (that is, services having the same labels), the service with higher weight is selected. This step ensures fair scheduling as it makes sure that the longer SFCs and the SFCs, which have waited longer in the queue get a fair chance for their scheduling.

Table II. Resource configuration taken from Amazon EC2.

| Name | API Name | Memory | Cores | Max Bandwidth | On Demand cost |
|---|---|---|---|---|---|
| T2 Small | t2.small | 2.0 GB | 1 cores | 25 MB/s | $0.034 hourly |
| T2 Medium | t2.medium | 4.0 GB | 2 cores | 25 MB/s | $0.068 hourly |
| T2 Large | t2.large | 8.0 GB | 2 cores | 25 MB/s | $0.136 hourly |
| M4 Large | m4.large | 8.0 GB | 2 cores | 56.25 MB/s | $0.140 hourly |

While selecting the VMs/PMs for service deployment, the affinity between services is taken into consideration. Two services belonging to the same instance of an SFC are considered to have higher affinity, and we try to place them on the same machine. This ensures minimum delays and less inter-machine traffic overhead. This step ensures that the services for the same SFC are scheduled on the same machine, if possible, to minimize the total traffic generated. Otherwise, it tries to schedule the service on the machine with which inter-machine traffic will be minimized, and all capacity constraints are satisfied. We may combine two or more services and deploy them on a single machine as well, provided a machine of that capacity is available. Availability of the machines depends on the cloud capacity. If a service instance is not serving any user demands, it is buffered in the cloud. In the buffered stage, the service uses fewer resources (such as storage only to save the state). However, it can be brought up quickly whenever relevant user demand arrives, saving resources and time [16]. For simplicity, we assume clouds have infinite buffering capacity. The steps for FWS algorithm are given in detail in Table III.

Table III. FWS algorithm for micro-service scheduling [49].

1. Let $\{1, 2, 3, 4, …, N\}$ be the set of micro-services to be scheduled on M machines. Let $\{T_1, T_2, T_3, …, T_n\}$ be their finish times.
2. If $T_i < T_j$ then MS j is said to be immediate successor of task i.
3. Let $S(i)$ be the set of all immediate successors of MS i.
4. Let $L_i$ be the label assigned to MS i.
5. Choose MS i from the arrived request s.t. $S(i) = 0$. Let $L_i$ be 1.
6. For $l = 2$ to N
7. Let C be the set of unlabeled MS s.t. there is no unlabeled successor.
8. Let s be the MS in C s.t. $T_s < T_{s*}$ for all other MS s* in C
9. Let $L_s = l$
10. Once labels are assigned, we assign weights $\{w_1, w_2, w_3, …, w_n\}$ to the services, s.t.:
$w_i \propto$ (number of dependent services in that chain) and
$w_i \propto$ (time spent by the services in the waiting queue).
11. Foreach service i
12. if $l_i = l_{i+1}$
13. select i for scheduling if $w_i > w_{i+1}$
14. else select $i+1$
15. Select the machine from the sorted list as per the remaining capacity for deployment.

We consider a 20-node topology out of which, 16 are the micro-clouds deployed at the edges such as cellular base stations, closer to the end users and four are core public clouds, with larger capacities, as shown in Fig. 14. Computation and/or data intensive services which need more processing and/or storage capacities and which tend to run for longer times, such as firewall, database services, are generally deployed at core clouds. We assume that each service instance produces data in the range of 5 kB to 20 kB. Also, the number of user requests each micro-service instance can handle at average load is selected from a range of 20 to 100 requests/sec. Time needed for execution of each service is chosen from the range of 10 to 100 milliseconds (ms) [18].

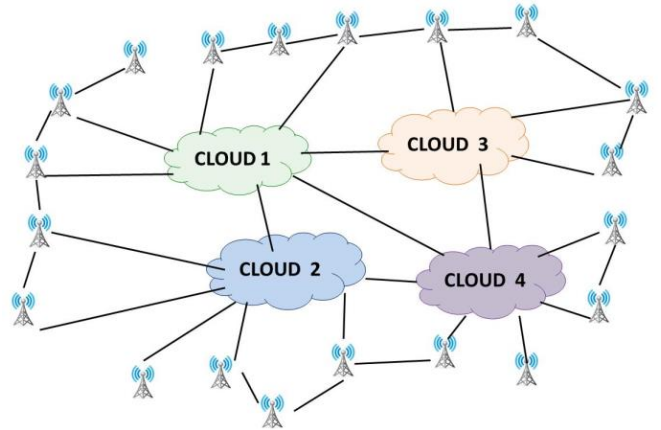

Fig. 14. 20-node topology with 16 micro-clouds and four core clouds.

All the values are selected randomly from the given ranges. In addition, we assign each user request with some delays and cost constraints it may tolerate. We also make sure these constraints are satisfied while scheduling the micro-services on the clouds. In the next section, we compare the results of proposed FWS solution with four variants of standard biased greedy scheduling strategy, which are common in the literature and observe significant improvements.

## VI. Results and Analysis

We now present the results obtained through the experimental setup. In addition to our FWS approach, we have implemented four additional algorithms based on the greedy biased approach for comparison. Service labeling step is common for all the heuristics. Table IV displays the basic steps for the following strategies:
1. Least-full First with First Finish (LFFF)
2. Most-full First with First Finish (MFFF)
3. Least-full First with Decreasing Time (LFDT)
4. Most-full First with Decreasing Time (MFDT)

Table IV. The selection criterion for greedy biased heuristics.

| Micro-Service | Machine Selection | |
| --- | --- | --- |
| | LF | MF |
| FF | Select Least full machine first, Select the service of SFC having first finish time | Select Most full machine first, Select the service of SFC having first finish time |
| DT | Select Least full machine first, Select the service of SFC with longest finish time | Select Most full machine first, Select the service of SFC with longest finish time |

We execute the algorithms for a certain number of times, and then we take an average. Graphs in Fig. 15 show the comparison of the four approaches mentioned above and our FWS approach in terms of the total inter-VM traffic generated. FWS approach (thick yellow line) performs the best with the least inter-VM traffic. For example, with 3000 user demands, greedy algorithms produce more than 20 MB of data, whereas FWS only produces less than 10 MB data, which is an improvement of 50%.

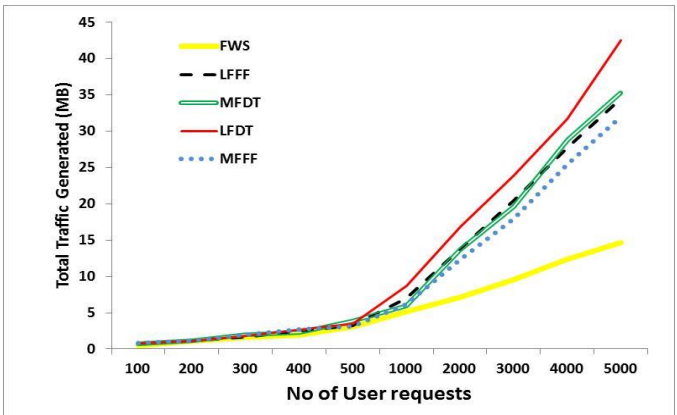

Fig. 15. Total traffic generated (in KB).

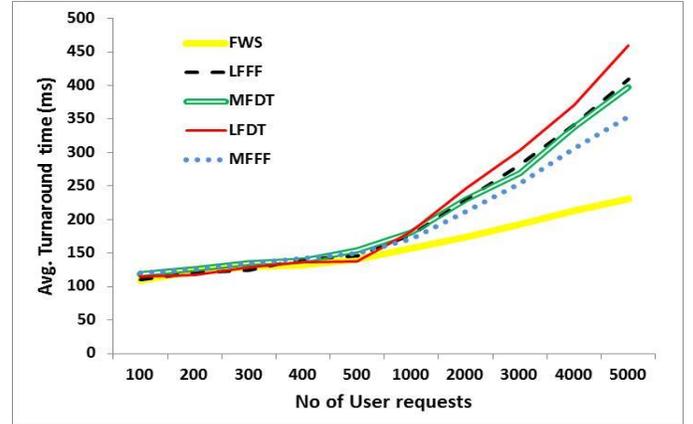

Fig. 16. Total turnaround time (in milliseconds).

Similarly, Fig. 16 shows the average turnaround time for each user where FWS again performs the best. For example, with 4000 user demands, FWS results in a turnaround time of less than 220 ms, whereas the other algorithms need around 330 ms We also present bar charts for above results, that is, for total traffic generated and average turnaround time in Fig. 17 and Fig. 18 respectively.

However, the average turnaround time alone is not sufficient to measure the performance, especially in the context of the time-sensitive applications. Most of the time, if the user demands are not satisfied within a given time constraint, it is as bad as service denied. Hence, we also find out the percentage of user demands which got satisfied in the given time constraints (Fig. 19). We observe that a significantly higher percentage of the user demands get satisfied with the FWS approach. The total percentage varies from 100% to 96% as user demands vary from 100 to 5000. On the contrary, the percentage drops to 70% for LFFF, 62% for LFDT & MFFF and 74% for MFDT.

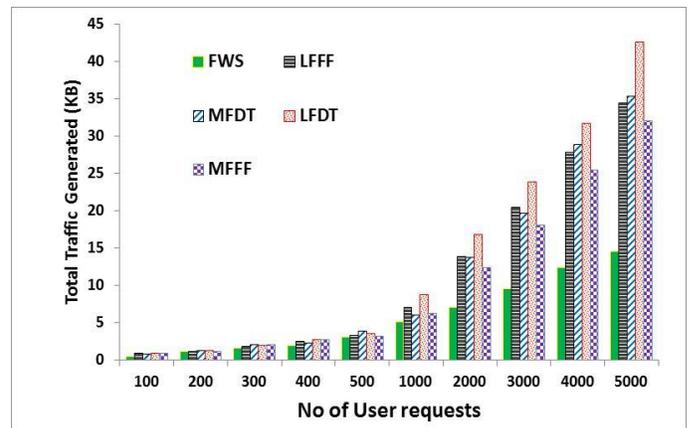

Fig. 17. Bar chart for total traffic generated (in KB).

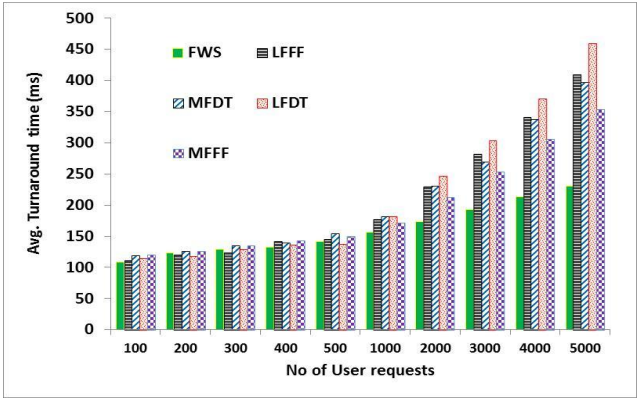

Fig. 18. Bar chart for total turnaround time (in milliseconds).

We have also analyzed the effect of traffic loads on average turnaround time or average time to schedule all the services. We observe exponential growth in the total turnaround delays as traffic loads in the network grow. We generated dummy traffic to obtain different average traffic loads. The links were modeled as M/D/1 queues, and by the standard formula, we calculate the delays in the links as given in Equation (1) below [23]. We note that $T_{ij}$ is the total delay on the link ($i$, $j$). $\lambda_{ij}$ is the arrival rate of packets and $\mu_{ij}$ is the processing rate of the same link. $T_{ij} = \frac{1}{2\mu_{ij}} \times \frac{2-(\lambda_{ij}/\mu_{ij})}{1-(\lambda_{ij}/\mu_{ij})}$ (1)

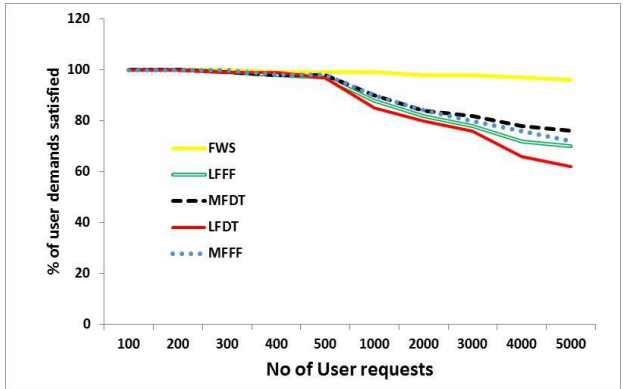

Fig. 19. Percentage of user demands satisfied.

In Fig. 20 we observe that even at 90% traffic load, the total delays, with the proposed FWS scheme remain within the range of 250 ms, which is within acceptable limits for the contemporary real-time applications [24]. For other schemes, however, it varies from 400 to more than 600 ms. In Fig. 21, we plot the graphs for the total costs of the resources needed to satisfy all the given demands using all the approaches. The cost has been calculated for an hour to host the required services for all the users.

We assume the Amazon pricing model as shown in Table II to calculate the costs. We observe that the proposed affinity-based FWS approach performs better than the greedy approaches in terms of the total cost as well. The cost difference goes on increasing with increase in the total number of users. This may be attributed to the fact that, in the affinity-based FWS approach, we try to accommodate the VMs, hosting micro-services with affinity, on a single machine with the closest match for the required capacities. This reduces the required number of the resources and eventually the cost. From the results, we observe that the proposed FWS scheme outperforms the contemporary greedy approaches in terms of the total traffic overhead, total turnaround time, the total number of the services satisfied as well as the total deployment cost.

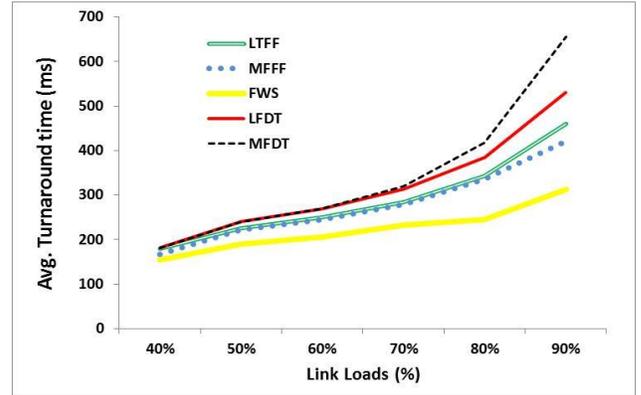

Fig. 20. Average turnaround time.

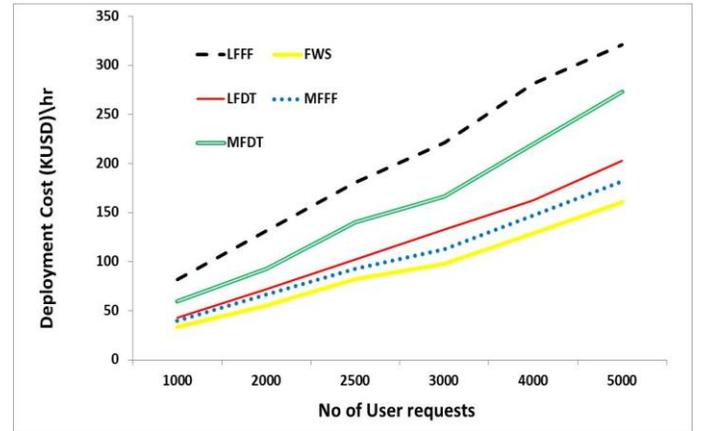

Fig. 21. Cost comparison (FWS vs. Greedy approaches).

## VII. Concluding Remarks and Future Work

In this paper, we discuss the micro-services and address important problems such as deployment and discovery of micro-services as well as communication among the different instances of the micro-services to form end-to-end service chains. In addition, we discuss the problem of scheduling micro-services. We point out that this is an important problem to be addressed for optimal service chains and point out the gap between the work done for virtual machine placement problem and micro-service scheduling problem. In addition, we point out that link loads and network delays while minimizing the total turnaround time and total traffic generated needs to be considered.

In this work, we aim to bridge the gap between the academia and the industry to help the service providers to deploy the micro-services more efficiently. In addition, we propose a novel FWS approach for micro-service scheduling in the multi-cloud scenario to form optimal SFCs. We take into account different delay and cost related SLAs. Also, we consider link loads and network delays while minimizing the total turnaround time and total traffic generated. The proposed approach demonstrates significant improvement compared to standard biased greedy approaches. However, there is still a wide area open for the research in developing novel scheduling algorithms considering the different delay and cost related SLAs. Advancements in the field of machine learning may be applied, such as proactive scheduling. Micro-service architecture brings in more challenges, such as distributed data management, failure recovery, security, monitoring, network latency, message formats, load balancing, fault tolerance and others, which need to be investigated further.

## Acknowledgment

This publication was made possible by the NPRP award [NPRP 8-634-1-131] from the Qatar National Research Fund (a member of The Qatar Foundation). The statements made herein are solely the responsibility of the author[s].